\documentclass[aps,prl,reprint,amsmath,amssymb,twocolumn,superscriptaddress]{revtex4-2}

\usepackage{graphicx} 
\usepackage{dcolumn}  
\usepackage{bm}       
\usepackage{xcolor}
\usepackage{graphicx}
\usepackage{subcaption}

\usepackage{booktabs}
\usepackage[normalem]{ulem}
\usepackage{color}
\definecolor{darkblue}{rgb}{0.0,0.0,0.4}

\usepackage{hyperref} 

\begin{document}

\title{Hadronisation of in-medium $c\bar c$ pairs to the exotic $X(3872)$} 

\author{Henrique Legoinha}
\email{h.legoinha@cern.ch}
\affiliation{Department of Physics, Tsinghua University, Beijing (THU), \\ \& Laboratório de Instrumentação e Física Experimental de Partículas (LIP)}

\author{Bernardo Picão}
\email{bernardo.picao@tecnico.ulisboa.pt}
\affiliation{Dep. de Física, Instituto Superior Técnico, Universidade de Lisboa (IST) \\ \& Centro de Física e Engenharia de Materiais Avançados (CeFEMA)}

\author{Pedro Bicudo}
\email{bicudo@tecnico.ulisboa.pt}
\affiliation{Dep. de Física, Instituto Superior Técnico, Universidade de Lisboa (IST) \\ \& Centro de Física e Engenharia de Materiais Avançados (CeFEMA)}

\begin{abstract} 
The production ratios of hidden-charm hadrons observed in nuclear collisions reflect the degree of separation that the precursor $c\bar c$ pair develops inside the quark-gluon plasma. The respective hadronisation amplitude manifests this relation through the position-space inner products of the in-medium and in-vacuum $c\bar c$ wave functions. The exotic $X(3872)$, whose structure accommodates a dominant loosely bound molecular component together with a smaller compact component, stands as a novel hard probe of this spatial dependence. 
In this Letter, relative production measurements of $J/\psi$, $\psi(2S)$ and $X(3872)$ hadrons performed by the LHC collaborations are understood through the spatial dispersion of in-medium charm anti-charm systems.
\end{abstract}

\maketitle


{\em Introduction -- } Ultra-relativistic heavy-ion collisions
create some of the most extreme conditions of temperature and density in the universe, in which ordinary
matter transitions into a deconfined, chirally symmetric
state~\cite{meltingH_boiling_quarks} known as Quark-Gluon Plasma (QGP). At the Large Hadron Collider (LHC), the QGP formed in Pb-Pb collisions exhibits a strong collective behaviour over femtoscopic length and time scales, rendering it a unique environment to test predictions from theory, namely quantum chromodynamics (QCD), both in its perturbative and non-perturbative regimes~\cite{annurev:/content/journals/10.1146/annurev-nucl-101917-020852,2025219,2024aliceqcdreview}. Comparisons with p-p collisions, where the produced system is much smaller, provide a baseline to understand the QGP properties and effects, with hard probes being an excellent phenomenological tool to register them~\cite{CMS:2026,APOLINARIO2022103990}.

Charm and anti-charm quark pairs are hard-scattered during the very early stages of nuclear collisions, much before the QGP reaches equilibrium~\cite{Schlichting:2019abc}. The dense color environment around them screens the QCD binding potential, weakening it, while interactions with roaming quanta drive the charm and anti-charm quarks to participate in the collective motion of the medium~\cite{ALICE:2021DmesonFlow,CMS:2018D0Flow}. These effects can leverage each other, dissociating hidden-charm systems into open-charm systems and so reducing the probability for charmonia production~\cite{Matsui:1986dk}. The $X(3872)$, an exotic hadron candidate, is extremely sensitive to such medium-induced effects~\cite{PhysRevLett.91.262001,PhysRevLett.106.212001,Braaten:2020iqw,ExHIC:2011say}. Its quantum numbers, $J^{PC}=1^{++}$, allow for an inner structure featuring an interplay between a dominant loose $D^0-\bar D^{0*}$ molecular state and a smaller compact state that accommodates possible tetraquark and charmonium-like contributions~\cite{Baru_2022,PhysRevLett.110.222001,Esposito_2025,LHCb:2024tpv,GUO2013127}. Hence, in-medium dissociated $c\bar c$ pairs can still evolve to this exotic hadron, giving its dominant molecular structure.

In this Letter, we take into account the QGP "melting" effects and connect the observed $X(3872)$, $\psi(2S)$ and $J/\psi$ production ratios at the LHC p-p and Pb-Pb collisions to the characteristic lengths that in-medium hidden-charm systems attain. This picture is well supported by Fermi's Golden Rule: the transition into hadrons should be proportional to the density of states, $\rho_{a}$ (with the subscript denoting the final hadronic state), and to the square of the inner product between the in-medium $c\bar c$ pair wave function (WF), $|\Psi_{c\bar c}^i\rangle$, and the in-vacuum hadronic WF, $|\Psi_{a}\rangle$. Then, hadron ratios follow from an overlap calculation given by: 
\begin{equation} 
\text{Had. ratios} \propto \frac{\rho_{a}}{\rho_{b}} \frac{|\langle \Psi_{a} |  \widetilde \Psi_{c\bar c}^i\rangle|^2}{|\langle \Psi_{b} | \widetilde \Psi_{c\bar c}^j\rangle|^2}\,,
\label{FERMIgr}
\end{equation}
where common normalizing factors cancel. Due to the large charm mass, the hadronisation of the $c\bar c$ pair can be described by a perturbative Hamiltonian (in the Born-Oppenheimer sense) which is then absorbed into the modified in-medium WF of the pair, $| \widetilde \Psi_{c\bar c}^i\rangle= H' | \Psi_{c\bar c}^i\rangle$.




{\em Experimental measurements --} The production of charmonia states as a function of their transverse momentum, $p_T$, and collision centrality, $Cent$, has been the subject of several experiments~\cite{Andronic:2015wma}. Measurements against centrality are particularly interesting since this variable (only available in Pb-Pb systems) conveys information about the size of the QGP, while measurements against $p_T$ can highlight the role of distinct production and hadronisation mechanisms. In turn, these form the baseline for building composite observables, namely the nuclear modification factor, $R_{AA}$, the relative production ratio, $R$, and its double ratio $DR$. 

The $R_{AA}$ quantifies the difference between a Pb-Pb and a binary-scaled p-p system in terms of particle counts. At the LHC, measurements of this observable using $J/\psi$ and $\psi(2S)$ mesons revealed an overall production suppression ($R_{AA}<1$) in Pb-Pb collisions, being particularly stronger for $\psi(2S)$~\cite{RELATIVE_PhysRevLett.118.162301}. Differentiated against centrality, a trend towards unity is visible for $J/\psi$ as collisions become more peripheral~\cite{raa_CMS:2016mah,ATLAS_jpsi_psi2s_pbpb_pp}. Against $p_T$, little to no dependence is found between $7.5 < p_T < 20~\text{GeV}$, after which a similar trend forms towards unity. For $\psi(2S)$, the corresponding differential measurements lie at lower $R_{AA}$ values~\cite{RAA_CMS:2017uuv}.

The relative production of charmonium states is sensitive to the dynamics of in-medium hidden-charm systems. Studies of this observable reveal substantially less $\psi(2S)$ production relative to $J/\psi$. In p-p collisions, this relative production increases with $p_T$ at mid-rapidity~\cite{RELP_ATLAS:2023qnh,PhysRevLett.114.191802,CMS_2012_RELP}, starting at $R^{\psi(2S)}_{pp}=0.28 \pm 0.03$ at $p_T\approx10~\text{GeV}$ through $R^{\psi(2S)}_{pp}=0.34 \pm 0.04$ at $p_T\approx45~\text{GeV}$. In Pb-Pb collisions, the corresponding measurement was performed at forward rapidity, revealing a mild $p_T$ dependence and none on centrality within the current uncertainties~\cite{2025_LHCb_pbpb_RELP,ALICE:2022jeh_RELP_cent}. 

In addition to conventional charmonium states, the $X(3872)$ relative production to $\psi(2S)$ has also been studied at the LHC in p-p~\cite{X_ATLAS:2016kwu,X_CMS:2013fpt, LHCb:2020sey}, p-Pb~\cite{PhysRevLett.132.242301} and Pb-Pb~\cite{PhysRevLett.128.032001} collisions. In the p-p system, a strong relative suppression of $X(3872)$ production is found, with flat $p_T$ dependence between $15 < p_T < 50~\text{GeV}$ at mid-rapidity. In contrast, in the Pb-Pb system, a central value compatible with a much smaller relative suppression was found at $15 < p_T < 50~\text{GeV}$ and $|y|<1.6$. The double ratio of $X(3872)$ to $\psi(2S)$ relative production in Pb-Pb and p-p collisions is $DR^X=10.2\pm6.8$, approximately computed at $15< p_T <50~\text{GeV}$ in mid-rapidity. 


To ensure consistency among the plethora of results introduced above, maintain proximity to the QGP thermodynamics, and suppress cold nuclear and quark coalescence effects~\cite{Acharya_2024,STAR_2019134917,PHENIX_PhysRevLett.98.232301,RAA_ALICE:2023,CMS:2017exb}, preference is given to measurements compatible with an effective transverse momentum of $ p^{eff}_T\approx10~\text{GeV}$ and taken at $|y|<1.6$. For measurements reported in intervals spanning a large low-to-high $p_T$ region the effective value is still reasonable. Particle production is predominantly taking place at the lower end of extended $p_T$ intervals, rendering ratios more sensitive to bins therein~\cite{Acharya_2023,ALICE:2019hno}.
Selected results differentiated against centrality are summarized in Table~\ref{tab:raa_DR_RHBTsummary}, and paired with a Hanbury Brown--Twiss (HBT) radii, $R_{HBT}$ ~\cite{Brown01071954,Lisa_2005,PhysRevC.93.024905}. Although $DR^X$ is obtained at slightly higher $p_T$, hidden-charm production ratios show small dependence on this variable within this $p_T$ region, rendering it a reliable reference in this context.


\begin{table}[t!]
\centering
\renewcommand{\arraystretch}{1.2}
\setlength{\tabcolsep}{4pt}
\begin{tabular}{ccccc}
\toprule
\textbf{Cent} (\%) & $\mathbf{R_{AA}^{J/\psi}}$ & $\mathbf{R_{AA}^{\psi(2S)}}$ & $\mathbf{R_{\mathbf{HBT}}^{\mathrm{PbPb}}}$~[fm] \\
\midrule
$0$--$10$  & $0.25\pm0.03$ & $<0.14$       & $6.0$ \\
$10$--$20$ & $0.33\pm0.03$ & $0.12\pm0.06$ & $5.4$ \\
$20$--$30$ & $0.40\pm0.04$ & $0.21\pm0.08$ & $4.7$ \\
$30$--$40$ & $0.46\pm0.05$ & $0.25\pm0.09$ & $4.1$ \\
$40$--$50$ & $0.62\pm0.07$ & $0.19\pm0.11$ & $3.7$ \\
\midrule
 $0$--$100$ & $0.34\pm0.02$ & $0.13\pm0.04$ & $5.0$ \\
\bottomrule
\end{tabular}
\caption{Summary of centrality-dependent nuclear modification factors and proxy Hanbury Brown--Twiss radii~\cite{RAA_CMS:2017uuv,RELATIVE_PhysRevLett.118.162301,PhysRevC.93.024905}. The differential entries correspond to $|y|<1.6$ and $6.5<p_T<30$~GeV/$c$; the inclusive entries correspond to $|y|<1.6$ and $9<p_T<12$~GeV/$c$. The $0$--$10\%$ $\psi(2S)$ $R_{AA}$ is a $95\%$ confidence-level upper limit.}
\label{tab:raa_DR_RHBTsummary}
\end{table}


\textit{Density of States -- } The density of states in Eq~\ref{FERMIgr} can be estimated using the Statistical Hadronisation Model~\cite{Andronic_2019,Braun_Munzinger_2004,andronic2026statisticalhadronizationsuccessesopen,Braun_Munzinger_2016}, where final-state hadrons are sampled from a thermal source at temperature $T\simeq158~\mathrm{MeV}$~\cite{Borsanyi_2020}. Hence, the Lorentz-invariant phase space measure of the final state hadron is weighted by a thermal factor which, in $(p_T,y,\phi)$ coordinates, reads 
\begin{align}
    \frac{d\rho_{a}}{dp_T}\Bigg|_{p^{eff}_T} = \frac{g_{a}}{8\pi^2} \, p^{eff}_T
\int_{-1.6}^{1.6} dy\,\,e^{\left(-m_T^{a}\cosh y\right)/T},
\label{eq:DOS_pTy}
\end{align}
where the azimuthal integral $\int d\phi = 2\pi$ was already performed,  $ m_T^{a}=(\,p_T^2+m_{a}^2)^{1/2}$ and $g_{a}$ is the hadron spin degeneracy.  For the particles under consideration, $g_{a}$ cancels when taking the ratios of the respective density of states, which becomes an average with respect to the thermal weight of the reference hadron, $\int_{-1.6}^{1.6}dy\, \exp(-(m_T^{b}\cosh y )/T)$,
\begin{align}
\frac{\rho_{a}}{\rho_{b}}= \left\langle e^{\left(-\Delta m_T^{ab}\cosh y\right)/T}
\right\rangle\,,
\label{eq:DOS_ratio}
\end{align}
with $\Delta m_T^{ab}=m_T^{a}-m_T^{b}$. When the same hadron appears on both sides of the ratio, as in $R_{AA}$, $\Delta m_T^{ab}$ vanishes and so ${\rho_{a}/\rho_{b}}$ reduces to unity. Similarly, in double ratios, the density of states ratio reduces to unity. Only for the $\psi(2S)$ to $J/\psi$ relative production a non-trivial density of states contribution, $\rho_{\psi(2S)}/\rho_{J/\psi}= 0.30$, appears.





{\em In-vacuum Wave Functions -- } To obtain $J/\psi$ and $\psi(2S)$ WFs we solve the Schrödinger equation for the potential
\begin{align}
    V_{c\bar c} =& -\frac{4}{3}\frac{\alpha_s}{r} + br 
    + \frac{32\pi}{9}\frac{\alpha_s}{m_c^2} \, \tilde{\delta}_\sigma(r) \,\vec S_c \cdot \vec S_c  \nonumber\\ 
    +&\frac{1}{m_c^2}\Bigg[\Bigg( \frac{2\alpha_s}{r^3} - \frac{b}{2r} \Bigg)\vec L\cdot\vec S + \frac{4\alpha_s}{r^3}T\Bigg], 
    \label{HNR}
\end{align}
treating all terms together under a coupled-channel framework to account for possible spin-tensor induced mixing between states sharing identical quantum numbers~\cite{Barnes:2005pb,Godfrey:1985xj}. The short-distance singularities arising from the spin-orbit and tensor interactions are regularized according to $1/r^3\;\longrightarrow\;\left(r^2+1/m_c^2\right)^{-3/2}$. The contribution of D-waves to the WFs of $J/\psi$ and $\psi(2S)$ are found to be smaller than $0.1\%$. Thus we consider their WFs to be correctly approximated by the $1S$ component.

\begin{figure*}[t!]
    \centering
    \includegraphics[width=0.95\linewidth]{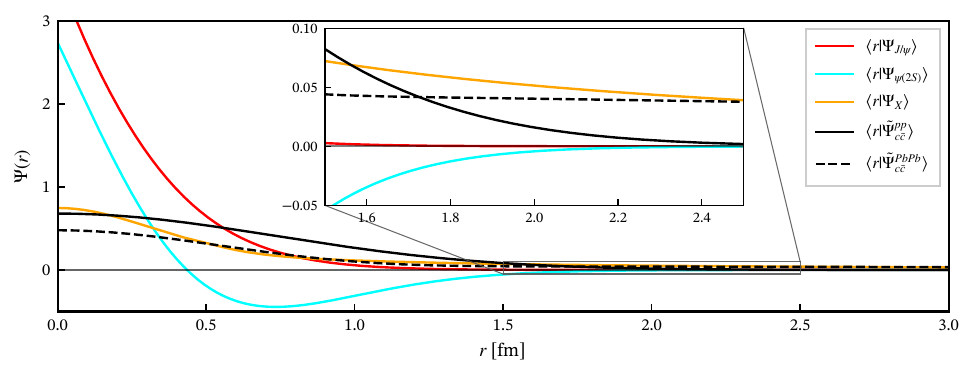}
    \caption{Effective in-medium $c\bar c$ modified WFs and in-vacuum states WFs. In this plot, the modified WF of the $c\bar c$ pair in the p-p collision system features $\sqrt{\langle r^2 \rangle}\approx0.9~\text{fm}$, while the narrow Pb–Pb component has $\sqrt{\langle r^2 \rangle}\approx0.65~\text{fm}$, with $\alpha=20\%$ such that the whole WF has $\sqrt{\langle r^2 \rangle}\approx4.5~\text{fm}$.}
    \label{fig:WFs}
\end{figure*}

For the exotic $X(3872)$ hadron, Low-Energy Universality (LEU)~\cite{Braaten:2003he,Braaten:2005jj} fixes the asymptotic form of the molecular component, $|\mathcal{M}(a)\rangle$. This is valid for shallow resonances with large scattering lengths, as is the case for $X(3872)$ at $a^X \approx9.6\pm1.7~\text{fm}$~\cite{JI20264416}. To also incorporate its compact nature, a Gaussian WF,
\begin{equation}
     \langle r| G(\sigma)\rangle = (\sqrt{\pi}\sigma)^{-3/2} \,\text{exp}({-r^2/(2\sigma^2)}) 
\end{equation}
is considered such that the full WF is a coherent sum,
\begin{equation}
    |\Psi_X\rangle = N\Bigg(\sqrt{\beta}\,|G(\sigma^X)\rangle + \sqrt{1-\beta}|\mathcal{M}(a^X)\rangle\Bigg),
    \label{X_compact_leu}
\end{equation}
with $N$ a normalizing factor, and with the LEU component regularized at short distances through a flat cap at $r=0.5~\text{fm}$. A value of $\beta=10\%$ is chosen as representative of the existing literature~\cite{JI20264416,Esposito_2025,Coito:2012vf,Hanhart_2011} and $\sigma^X$ is set so the compact WF features a $\sqrt{\langle r^2\rangle}\approx0.40~\text{fm}$, thus respecting the landscape of charmonia states, including hidden-charm tetraquarks~\cite{Yu_2024}. Still, due to its dominant delocalised nature, the $X(3872)$ WF remains very spatially extended, with $\sqrt{\langle r^2\rangle}\approx6.44~\text{fm}$.


{\em In-medium $c\bar c$ Wave Function -- } Owing to the stochastic nature of medium-induced dissociation, the modified WFs of $c\bar c$ pairs could be profiled through Gaussian ansätze constrained to respect the sizes of the p-p and Pb-Pb systems. These can be estimated using, as a proxy, the homogeneity lengths found in HBT measurements of the respective particle-emitting sources~\cite{PhysRevC.99.044913,PhysRevC.59.1585}. The perturbation Hamiltonian $H'$ is not expected to significantly modify the scale of the WFs. In Pb-Pb collisions, the measured radii, $R^{\mathrm{PbPb}}_{\mathrm{HBT}}$, are consistent with the formation of a large system, ranging up to $7~\text{fm}$ in the most central collisions~\cite{ALICE:2011dyt}. For other centralities, see Table~\ref{tab:raa_DR_RHBTsummary}. In contrast, in p-p collisions the system is much smaller, with $R^{\mathrm{pp}}_{\mathrm{HBT}}$ of the order of $1~\text{fm}$~\cite{ALICE:2011kmy}. Hence, these characteristic lengths provide a reference scale for the separation that dissociated $c\bar c$ pairs can develop in each system before hadronisation. Therefore, in the small p-p collision system, the $c\bar c$ pair modified WF is described by a single Gaussian,
\begin{equation}
 \langle r|\widetilde \Psi^{pp}_{c\bar c}(\sigma)\rangle=G(\sigma; r),
\end{equation}
with $\sigma$ the pair spatial width. Following the considerations above, this parameter is constrained such that it yields a modified WF featuring $\sqrt{\langle r^2\rangle}\leq R^{pp}_{\mathrm{HBT}}$. A lower limit can also be imposed at around $\sqrt{\langle r^2\rangle}\geq0.3~\text{fm}$, since smaller WFs do not fit well in the charmonia scale.

For the Pb-Pb system, the large spatial extent of the medium should allow to differentiate a compact component, associated with pairs that survive in a close configuration, from a more diffuse component, accounting for dissociated pairs whose separation increased over larger distances~\cite{Armesto:2026fit,Asakawa:2003re,Brambilla:2008cx,Larsen:2026qvs,Umeda:2002vr,Datta:2003ww,Karsch:2002wv,bicudo2008}. So, due to the presence of the medium, the modified WF of a $c\bar c$ pair in a Pb-Pb system is written as the incoherent superposition~\cite{PhysRevD.85.105011,PhysRevD.97.014003} between a compact and a broad Gaussian component,
\begin{align} 
\label{inmedium_PbPb} 
|\widetilde \Psi^{PbPb}_{c\bar c}(\sigma',\alpha)\rangle=N\Big(\sqrt{\alpha}\,|G(\sigma')\rangle+i\sqrt{1-\alpha}\,|G(\varsigma)\rangle\Big) 
\end{align}
with $\sigma'$ and $\varsigma$ respectively describing the spatial width of the $c\bar c$ pairs that remain bound and become dissociated, and $\alpha$ the relative balance between these two components. Again, $\sigma'$ is required to yield a contribution to the overall WF falling within charmonia scales, with $0.3\leq\sqrt{\langle r^2\rangle}\leq1~\text{fm}$, while $\varsigma$ is fixed such that the broad component features $\sqrt{\langle r^2\rangle}= R^{\mathrm{PbPb}}_{\mathrm{HBT}}$, matching the proximate characteristic size of the medium produced in Pb-Pb collisions. The fraction of in-medium $c\bar c$ quarks retaining a close configuration is also not completely free -- it must lie sufficiently away from unity and from zero, as these two limiting regions would contradict nuclear collision experiments~\cite{annurev:/content/journals/10.1146/annurev-nucl-101918-023806}. Hence, $\sigma$, $\alpha$ and $\sigma'$ are allowed to float within a physically motivated range, to be tested against the experimental measurements of charmonia relative production in p-p and Pb-Pb collision. 

In Fig.~\ref{fig:WFs}, in-medium modified WFs of $c\bar c$ pairs in p-p and Pb-Pb systems are sketched, using representative parameters, alongside $J/\psi$, $\psi(2S)$ and $X(3872)$ WFs. The zoomed panel highlights the evanescent tail that the WF of $c\bar c$ pairs in a Pb-Pb system develops. It becomes very similar in shape to the $X(3872)$ WF.

\begin{figure}[t!]
    \centering
    \includegraphics[width=\linewidth]{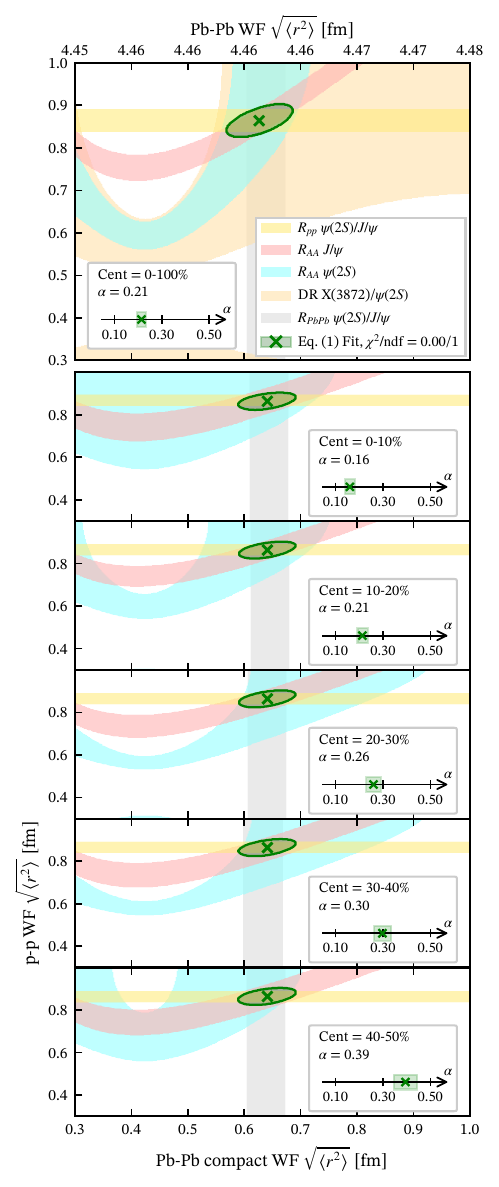}
    \caption{The 2D plots for several centrality bins as a function of the $\sqrt{\langle r^2\rangle}$ of $c \bar c$  in the Pb-Pb and p-p media. For the inclusive case (top plot), the total $\sqrt{\langle r^2\rangle}$ of the $c\bar c$ in the Pb-Pb medium is shown in the top axis. The forward rapidity $R_{\mathrm{PbPb}}^{\psi(2S)}$, shaded in grey and not considered as a fit constraint, is also shown.}
    \label{fig:2Dmaps_Cent}
\end{figure}

\textit{Results -- } Through Eq.~\ref{FERMIgr}, the best-fit values of $(\sigma,\sigma',\alpha)$ are determined by minimizing $\chi^2$ against the measurements introduced above. The widths, $\sigma$ and $\sigma'$, are shared across the five centrality bins, while the compact fraction, $\alpha$, is passed independently. The inclusive centrality case is fitted separately. Figure~\ref{fig:2Dmaps_Cent} outlines the experimentally allowed phase space at the best-fit value of $\alpha$ in the $(\sigma,\sigma')$ plane, expressed through the corresponding $\sqrt{\langle r^2\rangle}$. The bands are overlaid with the fit minima and corresponding $1\sigma$ regions. 

A single common overlapping region is found. It remains stable across the centrality bins and evolves smoothly in parameter space. The permitted values of $\alpha$ support the scenario in which a fraction of the $c\bar c$ pairs produced in Pb-Pb collisions remain sufficiently localized, with $\sqrt{\langle r^2\rangle}=0.63^{+0.05}_{-0.05}$~fm, to project efficiently onto charmonia states. This surviving fraction increases towards peripheral collisions, from $\alpha=0.16^{+0.02}_{-0.02}$ through $\alpha=0.39^{+0.05}_{-0.05}$, encoding the centrality dependence of the experimental measurements. In each medium, the characteristic separation of a $c\bar c$ pair is $\sqrt{\langle r^2\rangle}=0.86^{+0.03}_{-0.03}$~fm in the p-p system and $\sqrt{\langle r^2\rangle}=4.46^{+0.05}_{-0.05}$~fm for the Pb-Pb system. The plot at the top of Fig.~\ref{fig:2Dmaps_Cent} shows the centrality-inclusive case, with the $X(3872)$ experimental constraint successfully accommodated. The fitted parameters suggest $X(3872)$ has $R_{AA}=1.39^{+0.92}_{-0.23}$, consistent with the current reasoning~\cite{PhysRevLett.128.032001,PhysRevC.110.L021901,PhysRevC.105.054901}. Systematic studies can be inspected in the supplemental material.

\textit{Conclusion -- } In this Letter, we use measurements of $J/\psi, \, \psi(2S)$ and $X(3872)$ production in nuclear collisions to establish a common hidden-charm hadronisation framework. The strong overlap between $|\widetilde \Psi^{PbPb}_{c\bar c}\rangle$ and the LEU component of $|\Psi_X\rangle$ drives $X(3872)$ yield in the Pb-Pb system, exposing the presence of spatially extended $c\bar c$ configurations. Our work highlights how sensitive hadronisation is to the spatial distribution of the involved states. In this context, the Golden Rule is a bridge between hadron theory and heavy-ion phenomenology for resolving heavy quark structures.

Future centrality-dependent measurements of spatially extended hidden-charm exotics will provide essential constraints on the separation reached by in-medium $c\bar c$ pairs. Employing the $X(3872)$, the width of the dissociated $c\bar c$ component can be directly fitted within the present framework.
A complementary study with the \(T_{cc}^{+}\) would provide insights about the spatial evolution of a $cc$ pair instead~\cite{2022lhcb_tcc}. Purely compact states like hidden-charm tetraquarks~\cite{LHCb:2018oeg,PhysRevLett.112.222002} and pentaquarks~\cite{PhysRevLett.122.222001,Aaij_2023,Aaij_2015,PhysRevLett.128.062001}, including with strangeness~\cite{LHCb:2021uow,Chatrchyan_2014,Aaij_2017,ALICE:nphys2017,PhysRevC.77.044908}, or, in the beauty sector, the $\Upsilon(nS)$ family~\cite{CMS:2018zza,Khachatryan_2017,Sirunyan_2018}, can further constrain the surviving compact arrangement while testing theoretical bound-state model predictions. 



\clearpage

\textit{Acknowledgments -- } 
This work was partly supported by TSU, Tsinghua University of Beijing, under the Tsinghua University Initiative Scientific Research Program; and by LIP, Laboratório de Instrumentação e Física Experimental de Partículas, under the contract UID/50007/2025 with FCT, Fundação para a Ciência e Tecnologia; and by CeFEMA, Center of Physics and Engineering of Advanced Materials, under contracts UID/04540/2025, UID/PRR/04540/2025 and UID/PRR2/04540/2025 with the Agência para a Investigação e Inovação AI2, https://doi.org/10.54499/UID/PRR/04540/2025, https://doi.org/10.54499/UID/PRR/04540/2025 and https://doi.org/10.54499/UID/PRR2/04540/2025 ; and by LaPMET, Laboratory of Physics for Materials and Emerging Technologies under contract LA/P/0095/2020 with the Agência para a Investigação e Inovação AI2, https://doi.org/10.54499/LA/P/0095/2020.

\bibliography{biblio}

\begin{figure}[b!]
    \centering
    \includegraphics[width=0.92\linewidth]{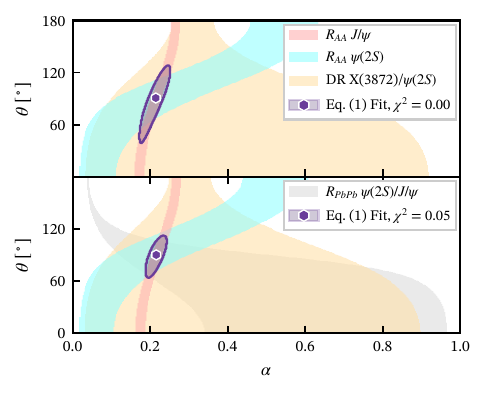}
    \includegraphics[width=0.94\linewidth]{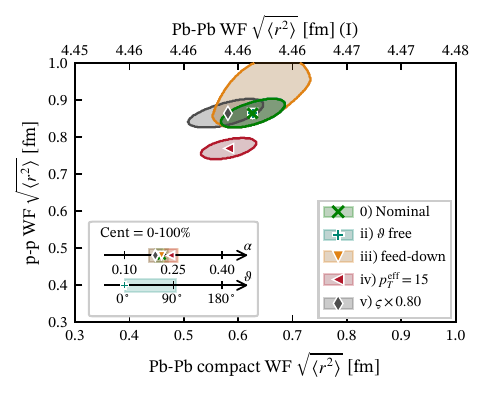}
    \includegraphics[width=0.92\linewidth]{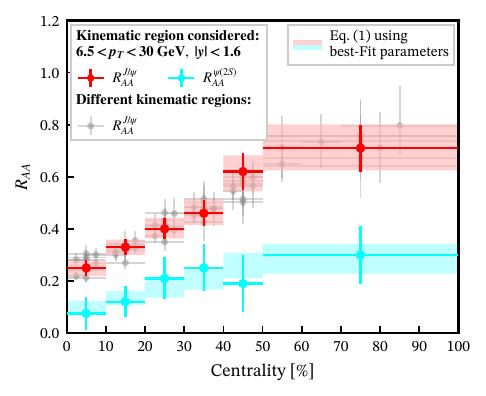}
\caption{Systematic variations of the nominal framework in the inclusive $0-100\%$ bin (top and middle panels). Quoted measurements of $J/\psi$ (red) and $\psi(2S)$ (cyan) nuclear modification factor as a function of centrality compared to Eq.~\ref{FERMIgr} evaluated at the nominal best-fit parameters. The grey points feature $J/\psi$ $R_{AA}$ in different $p_T$ and $|y|$ regions. The associated experimental uncertainties have been combined~\cite{RAA_CMS:2017uuv,raa_CMS:2016mah}.}
\label{fig:syst_all}
\end{figure}

\appendix
\textit{Supplemental Material -- } To assess the robustness of the phenomenological study carried out in this Letter, we employ several variations of the considered nominal framework and refit the data constraints. Namely, these variations involve i) setting free the relative phase, $\theta$, between compact and dissociated $c\bar c$ states in the Pb-Pb system; ii) setting free the relative phase, $\vartheta$, between the loose and compact components entering the $X(3872)$ WF; iii) possible feed-down contaminations in the charmonia ratios~\cite{Faccioli:2008ir}; iv) DoS sensitivity with $p^{eff}_T$; and v) sensitivity to a smaller $c\bar c$ separation. In the upper panel of Fig.~\ref{fig:syst_all}, the $(\theta,\alpha)$ plane is plotted at the best fit $\alpha$. In the lower subplot, we add the forward rapidity measurement of $\psi(2S)$ relative production against $J/\psi$ in Pb-Pb collisions to the fit constraints.

i) The refit returns $\theta=90.8^{\circ+37.5}_{-51.8}$ and, if the additional constraint is considered, $\theta=90.0^{\circ+22.2}_{-26.2}$, which further emphasizes the destructive interference. Hence, an incoherent sum between the compact and broad components of the WF of $c\bar c$ pairs in the Pb-Pb system is favoured. This is expected since the presence of the stochastic medium should average the relative phase away.

In the middle panel, the new fit minima for the centrality-inclusive case are shown at the best $\alpha$ in the $(\sigma,\sigma')$ plane, expressed via the corresponding $\sqrt{\langle r^2\rangle}$.

ii) In the middle panel, we note that the $X(3872)$ WF has a strong preference for retaining a coherent sum of the contributing states. The best fit returned $\vartheta=0^{\circ+95}_{-0}$, which corroborates a plain coherent sum. Within the current experimental uncertainty, the value of $\beta$ cannot yet be constrained, therefore a median value from literature was used. Further measurements, especially centrality-dependent ones, will allow to pinpoint this value.

iii) We note that a $30\%$ feed-down contamination in the charmonia ratios enlarges the uncertainty of the in-medium $c\bar c$ separation. It does not spoil the simultaneous treatment of the hidden-charm states considered.

iv) We see that a significant increase in $p^{eff}_T$ shifts the overlap region slightly downwards, allowing the overall conclusions to be maintained. We thus expect this DoS factorization to be valid for a range $5<p_T<15$ GeV.

v) Considering a $20\%$ smaller separation of the dissociated $c\bar c$ component in the Pb-Pb system still yields a healthy common treatment of these particles. Further trimming the separation of the dissociated $c\bar c$ pair returns unphysical parameters (very compact $c\bar c$ configurations in a Pb-Pb environment). Excluding the Gaussian ansatz that accounts for the spreading $c\bar c$ pair prevents the fit from converging. 

The bottom panel shows how the best parameters found when fitting the selected data with the nominal setup relate to measurements taken at different rapidity intervals, including $|y|>1.6$, $|y|<2.4$, $|y|<1.2$ and $|y|<0.6$, and at larger $p_T$, up to $50~\text{GeV}$.

\end{document}